\documentstyle[eqsecnum,epsfig,twocolumn,aps,floats,graphicx]{revtex}
\begin{document}
\draft \preprint{}

\twocolumn[\hsize\textwidth\columnwidth\hsize\csname
@twocolumnfalse\endcsname

\title{Bilayer Splitting in the Electronic Structure of Heavily
 Overdoped Bi$_2$Sr$_2$CaCu$_2$O$_{8+\delta}$}
\author{D. L. Feng, N. P. Armitage,
D. H. Lu,  A. Damascelli, J. P. Hu, P. Bogdanov, A. Lanzara, F.
Ronning, K. M. Shen, H. Eisaki, C. Kim, and Z.-X.Shen}
\address{
Department of Physics, Applied Physics and Stanford Synchrotron
Radiation Laboratory,\\ Stanford University, Stanford, CA 94305,
USA }
\author{J.-i. Shimoyama, and K. Kishio}
\address{
Department of Applied Chemistry, University of Tokyo, Tokyo,
113-8656, Japan}

\date{Jan. 17, 2000}
\maketitle

\begin{abstract}

The electronic structure of heavily overdoped
Bi$_2$Sr$_2$CaCu$_2$O$_{8+\delta}$  is investigated by
angle-resolved photoemission spectroscopy. The long-sought bilayer
band splitting in this two-plane system is observed in both normal
and superconducting states, which qualitatively agrees with the
bilayer Hubbard model calculations. The maximum bilayer energy
splitting is about 88 meV for the normal state feature, while it
is only about 20 meV for the superconducting peak. This anomalous
behavior cannot be reconciled with the quasiparticle picture.

\end{abstract}

\vskip2pc ]

\narrowtext

High temperature superconductors (HTSC's), as doped Mott
insulators, show strong doping dependent behavior. The underdoped
regime of the HTSC's is characterized by its unconventional
properties, such as the pseudogap and non-Fermi liquid transport
behavior. On the other hand, the overdoped regime is considered to
be more ``normal", partly because of the absence of a pseudogap
and more Fermi liquid-like behavior. It is very challenging and
important for HTSC theories to be able to explain the
phenomenology in both regimes. Angle resolved photoemission
spectroscopy (ARPES), one of the most direct probes of the
electronic structure, has contributed greatly to the understanding
of the electronic structure of the HTSC's. However, most systems
studied by ARPES have either low $T_c$'s (below 40K for
La$_{2-x}$Sr$_x$CuO$_{4+\delta}$ (LSCO), and
Bi$_2$Sr$_2$CuO$_{6+\delta}$ (Bi2201)), or doping limitations
(only up to slightly overdoping for
Bi$_2$Sr$_2$CaCu$_2$O$_{8+\delta}$ (Bi2212) and
YBa$_2$Cu$_3$O$_{7-y}$ (YBCO)). For a complete understanding, it
is very important to study the heavily overdoped systems,
especially Bi2212, which is the most studied system by ARPES.

Recent advances in high pressure annealing techniques have made it
possible to synthesize heavily overdoped Bi2212. In this paper, we
report ARPES measurements of the electronic structure of heavily
overdoped Bi2212. We show that the long-sought bilayer band
splitting (BBS) exists for both normal and superconducting states
of this material over large fraction of the Brillouin zone. The
detection of the BBS, which has been predicted by band structure
calculations\cite{LDA,OKAnderson95}, but not observed in earlier
ARPES data\cite{Ding96}, enables us to address several important
issues. First, it provides a very detailed test for the
theoretical calculations, with our experimental results favoring
the bilayer Hubbard model\cite{OKAnderson96} over LDA
calculations\cite{LDA,OKAnderson95}. Second, it shows the novel
result that the bilayer splitting energy in the superconducting
state is only about 23\% of the normal state splitting. Third, it
provides an explanation for the detection of a ``peak-dip-hump"
structure in the normal state of heavily overdoped
samples\cite{Rast00,WellsUP}.

Heavily overdoped Bi2212 samples ($T_C (onset)=65$ K, $\Delta T_C
(10\%\sim90\%) = 3$ K, denoted as OD65) were synthesized by
annealing floating-zone-grown single crystals under oxygen
pressure $P_{O_2}=300$ $atm$ at 300$^\circ$C for two weeks, and
characterized by various techniques. Magnetic susceptibility
measurements show that the presence of a second phase is less than
1\%. Laue diffraction and low energy electron diffraction (LEED)
patterns show that its superstructure and surface resemble those
of optimally doped samples, and the flatness of the cleaved sample
surface is shown by the small laser reflection from the sample.
Angle resolved photoemission experiments were performed at
beamline V-4 of Stanford Synchrotron Radiation Laboratory (SSRL)
with a Scienta SES200 electron analyzer, which can take spectra in
a narrow cut of $0.5^\circ\times14^\circ$ simultaneously in its
angular mode with an angular resolution as good as  $0.12^\circ$
along the cut direction. The data were collected with polarized
synchrotron light from a normal incidence monochromator, where the
intensity of the second order light is extremely weak, as well as
He-I light from a He discharge lamp. The overall energy resolution
varied from 10 meV to 17 meV under different conditions. The
chamber base pressure was better than $5\times10^{-11} torr$
during the measurements.

ARPES spectra were taken over a wide region of the Brillouin zone
of OD65. Fig. \ \ref{dispersion}a-h show the normal state
photoemission intensity as a function of momentum and binding
energy in false color. In this way, one can clearly see the
centroids of the dispersing features. For example, Fig.\
\ref{dispersion}a shows that one band disperses and crosses the
Fermi energy along a momentum cut that goes through the $d$-wave
node region. Away from the nodal region, this seemingly single
feature splits into two features, Features A and B, starting from
Fig.\ \ref{dispersion}c. The photoemission intensities in the
bracketed region are replotted in the form of energy distribution
curves (EDC's) in Fig.\ \ref{dispersion}i, where spectra show that
two peaks that cross the Fermi level less than 0.6$^\circ$ apart.
This splitting increases when approaching the $(\pi,0)$ region. In
Fig.\ \ref{dispersion}h, the Features A and B are well-separated,
and two more weaker features are clearly visible as well; these
are the superstructure images of Features A and B, which are
typically about $(0.22\pi,0.22\pi)$ away from their corresponding
main features in Bi2212. The absence of splitting in the nodal
region is checked with the best achievable angular resolution
($\sim0.12^\circ$).

The observed two bands in the normal state spectra of Bi2212
samples can be naturally interpreted by the presence of the BBS.
Because the Bi2212 ARPES features are considered to be mainly
contributed by the anti-bonding $x^2-y^2$ state in the CuO$_2$
plane, and Bi2212 has two CuO$_2$ planes per unit cell, the
intrabilayer coupling would cause a splitting. As we will see
later, the observed splitting agrees with what is expected from a
bilayer system\cite{OKAnderson96}. This interpretation is also
supported by recent studies of heavily overdoped single-layer
Bi2201, where only one band was observed\cite{FengUP}. Since
Feature A is always at lower binding energy than Feature B at a
given momentum, we assign the anti-bonding band (AB) to Feature A,
and bonding band (BB) to Feature B.

The Fermi surfaces (FS's) can be determined by determining Fermi
crossings of the bands (dispersion method), or determining the
local maxima of the low energy ARPES spectral weight distribution
(spectral weight method) (Fig.\ \ref{fs})\cite{method}. One can
see two main FS's, one for the antibonding band (AB) and the other
for the bonding band (BB), and their corresponding superstructure
images (AB' and BB'). The observed hole-like Fermi surface
topology is consistent with early findings in less overdoped
Bi2212 systems at similar photon energies. These FS's overlap in
the nodal region and gradually depart from each other when
approaching the $(\pi,0)$ region. Fig.\ \ref{fs}b shows EDC's
along one cut that crosses all of the four Fermi surfaces. At 22.7
eV photon energy (lower right half of Fig.\ \ref{fs}a), the AB has
more weight near $E_F$ than the BB, and this situation is reversed
at 20 eV (upper left half of Fig.\ \ref{fs}a). This strong photon
energy dependence of the relative intensities of the AB and BB is
consistent with the BBS, because the AB and BB have odd and even
symmetries respectively along the c-axis. By tuning the incident
photon energy,  the wavevector of the final electron state along
the c-axis is changed, which changes the photoemission
cross-sections between the final state and the initial BB and AB
differently due to their opposite symmetries. The fact that we see
BBS all over the FS and in the superstructure images (AB' and BB')
away from the $(\pi,0)$ region rules out the possibility that the
split FS's are artifacts caused by the superstructure. Moreover,
because the intensity of BB is weaker than that of AB in the 22.7
eV photon energy data, AB cannot be a superstructure of BB, and
vice versa for the data taken at 20 eV photon energy.

To understand the effect of the BBS on the superconducting state,
spectra were taken in both the normal and superconducting states
near the $(0,\pi)$ region (Fig.\ \ref{pi0edc}), where the
splitting is greatest. Moreover, it was found that in this region,
the ARPES lineshape of Bi2212 evolves dramatically across $T_c$
from a broad spectrum in the normal state into a well-known
peak-dip-hump (PDH) structure in the superconducting
state\cite{Dessau91}.

In the normal state (Fig.\ \ref{pi0edc}a), the antibonding state
crosses $E_F$ near n4 and n-4, while the bonding state disperses
through the Fermi energy around spectra n8 and n-8. The presence
of two features in the normal state was reported earlier
\cite{Rast00,WellsUP}, and suggested to be an anomalous normal
state counterpart to the conventional superconducting
PDH\cite{Rast00}. Here, we show that this feature is actually due
to the bilayer splitting. In spectra n-3 through n3, the BB is at
high binding energy and thus broad, while the AB is at low binding
energy and thus sharp, which conspire to give a PDH-like
structure. We stress that this is fundamentally different from the
PDH structure that turns on at $T_c$.

In the superconducting state (Fig.\ \ref{pi0edc}b), the low energy
part of the spectra evolves into two sharp superconducting peaks.
It appears that both the normal state BB and AB develop their own
superconducting PDH structure. Similar to the superconducting peak
reported before in less overdoped samples, both BB and AB
superconducting peaks lose their intensity upon crossing the
corresponding normal state BB/AB FS's. More specifically, spectra
s7 and n7 (replotted in Fig.\ \ref{pi0edc}c), which consist mainly
of the BB, strongly resemble the normal and superconducting state
spectra from overdoped samples with less carrier
doping\cite{Feng00}. When the BB superconducting peak disperses to
higher binding energies, it becomes weaker and presumably
contributes very little to the sharp peak seen at s0. Therefore,
the observed sharp peak at s0 can be regarded as mainly due to the
antibonding state. For spectra containing two peaks, they can be
fitted by two PDH's, as shown in Fig.\ \ref{pi0edc}c for s4.

The dispersions extracted from Fig.\ \ref{pi0edc} are summarized
in Fig.\ \ref{energy}a. Because the superconducting peak intensity
of the BB is very weak near $(\pi,0)$, its position is
extrapolated and shown as the dotted line. Although the BB and AB
superconducting peaks have different dispersions, their minimum
binding energies near their respective FS's are almost the same
($\sim$16 meV), which shows that the BB and AB have the same
$d$-wave superconducting gap amplitude. The maximum energy
splittings can be extracted from the binding energies at
$(\pi,0)$. They are found to be about 88 meV for the normal state
bands, and interestingly, only about 20 meV for the
superconducting peaks. The striking difference in the splitting
energies cannot be explained conventional theories, where
quasiparticles below $T_c$ have an energy of $E_{\bf
k}=\sqrt{\Delta_{\bf k}^2+\varepsilon_{\bf k}^2}$, with
$\varepsilon_{\bf k}$ and $\Delta_{\bf k}$ being the normal state
quasiparticle energy and superconducting gap, respectively. The
small splitting energy of the superconducting peak also counters
the naive expectation that global phase coherence below $T_c$ will
enhance the $c$-axis coupling and thus cause larger splitting.
Instead, the data demonstrate a qualitative breakdown of this
quasiparticle concept. This conclusion is in concert with the
earlier observation that the weight of the superconducting peak is
closely related to the carrier doping level and the condensation
fraction of the system\cite{Feng00}. We hope the new data can
stimulate more theoretical works on this issue.

The nature of the normal state BBS as a function of momentum and
energy puts strong constraint on theoretical models. A maximum
momentum splitting near $(\pi,0)$ contradicts early LDA
calculations, where the calculated BiO Fermi surface near
$(\pi,0)$ causes very a small splitting of the CuO$_2$ bands near
$(\pi,0)$\cite{LDA}. However, it does agree with bilayer LDA
calculations that only consider bands from the two CuO$_2$
planes\cite{OKAnderson95}, and the bilayer Hubbard model, which is
based on the bilayer LDA band calculations plus additional on-site
Coulomb repulsion\cite{OKAnderson96}. The bilayer Hubbard model
predicts two AB/BB Fermi surfaces similar to the data for similar
carrier doping levels\cite{OKAnderson96}.

The bilayer LDA calculations \cite{OKAnderson95} predicted that
the normal state bilayer energy splitting to be $2t_\bot({\bf
k})=t_\bot(\cos(k_xa)-\cos(k_ya))^2/2$, where $t_\bot({\bf k})$ is
the anisotropic intrabilayer hopping. It indicates that the
maximum energy splitting is $2t_\bot$ at $(\pi,0)$. This agrees
with the data, and one obtains the experimental intrabilayer
hopping $t_{\bot,exp}=44\pm5$ meV. To test this over a large
momentum range, the normal state energy splitting along the AB
Fermi surface (Fig.\ \ref{energy}b) were extracted from the data
in Fig.'s\ \ref{dispersion} and \ref{fs}. Indeed, the data can be
fitted very well by $t_{\bot,exp} (\cos(k_xa)-\cos(k_ya))^2/2$,
but quantitatively, the experimental maximum energy splitting of
88 meV ($2t_{\bot,exp}$), is much smaller than the 300 meV
($2t_{\bot,LDA}$) splitting predicted by the bilayer LDA
calculations\cite{OKAnderson95}. On the other hand, the data agree
better with the bilayer Hubbard model\cite{OKAnderson96}, which
predicted a similar anisotropic energy splitting with 40 meV
maximum energy splitting at $(\pi,0)$ for the similar doping
level. This is because unlike the bilayer LDA calculations, the
bilayer Hubbard model considers strong correlations, and strong
on-site Coulomb repulsion (or correlations) will substantially
reduce the hopping to an occupied site thus reducing the effective
intrabilayer hopping. Based on this, its small splitting energy
scale (40 meV) may suggest that weaker on-site Coulomb repulsion
should be adopted in the bilayer Hubbard model (at least for the
heavily overdoped case). We note that $t_{\bot,exp}$ is of similar
magnitude of the gap, and is a significant fraction of the
in-plane exchange coupling $J$, and the bandwidth. Therefore, the
intrabilayer coupling should be considered in models describing
Bi2212.

A natural question is why the bilayer band splitting is
particularly prominent in heavily overdoped materials. This is
mainly because the more Fermi liquid-like behavior in the heavily
overdoped regime results in much better defined quasiparticles,
i.e., much sharper features. The absence of two well-defined
features in the spectra of less overdoped samples does not
necessarily imply the absence of the BBS. In fact, with improved
resolution, preliminary studies have found signatures of BBS in
the normal state of slightly overdoped Bi2212 samples\cite{BBSUP}.

In summary, the electronic structure of heavily overdoped
Bi$_2$Sr$_2$CaCu$_2$O$_{8+\delta}$  is investigated by
angle-resolved photoemission spectroscopy. The bilayer band
splitting in this two-plane system is observed in both normal and
superconducting states, which qualitatively agrees with the
bilayer Hubbard model calculations. The different energy splitting
scales reported here provide new information for the behavior of
the superconducting peak, which cannot be well understood in the
quasiparticle framework and needs further investigation.

Stanford Synchrotron Radiation laboratory is operated by the DOE
Office of Basic Energy Science Divisions of Chemical Sciences and
Material Sciences. The Material Sciences Division also provided
support for the work. The Stanford experiments are also supported
by the NSF grant 9705210 and ONR grant N00014-98-1-0195-A00002.

\begin{figure}[tbhp]
\caption{(color) (a-h) false color scale plot of the OD65 normal
state (T=75K) ARPES spectra taken with 22.7 eV synchrotron light,
they are along the momentum cuts indicated by lines in the inset.
Feature A and B, and their superstructure image A' and B' are
indicated by triangles, circles, squares, diamonds respectively.
The EDC's near the Fermi crossing  in (c) (indicated by ``$[$")
are plotted  in (i). The angular resolution is $0.3^\circ$.}
 \label{dispersion}
\end{figure}

\begin{figure}[tbhp]
\caption{(color) (a) False color plot of the momentum distribution
of the spectral weight near $E_F$ ([-20 meV, 10 meV]) of OD65
taken at 22.7 eV (lower right half, T=75K) and 20 eV (upper left
half, T=80K) (note they are from different experiments). The Fermi
surface determined by dispersion is also plotted for antibonding
states (AB, triangles), bonding states (BB, circles),
superstructure images of antibonding states (AB', squares), and
bonding states (BB', diamonds). (b) ARPES spectra along the cut
indicated by the arrow in (a).
 }
 \label{fs}
\end{figure}

\begin{figure}[tbhp]
\caption{ARPES spectra taken on OD65 with He-I light for (a)
normal state, and (b) superconducting state, where the
superconducting peak of the antibonding and bonding states are
indicated by crosses and bars respectively. The angular resolution
is $0.56^\circ$. (c) shows selected spectra from (a) and (b). Note
that the fit of s4 is not unique. The spectra are taken along
$(-0.24\pi,\pi)-(0.24\pi,\pi)$, and labeled from -9 to 9 as shown
in the inset of (c).}
 \label{pi0edc}
\end{figure}

\begin{figure}[tbhp]
\caption{(a) Dispersion extracted from Fig.\ \ref{pi0edc} of the
superconducting peaks for the bonding sate (bars) and the
antibonding state (crosses), and the normal state bands of the
bonding sate (solid circles) and the antibonding state
(triangles). (b) Energy splitting along the AB Fermi surface,
which are obtained from data shown in Fig.\ \ref{dispersion}. It
is simply the binding energy of the BB, since the binding energy
of AB is zero at its Fermi surface. The curve is $t_{\bot,exp}
[\cos(k_xa)-\cos(k_ya)]^2/2$, where $t_{\bot,exp}=44\pm5$ meV.
Error bars are due to the uncertainties in determining the energy
position.} \label{energy}
\end{figure}


\begin{references}
\bibitem{method} We note that at 20 eV photon energy, the FS's
determined by the dispersion method do not agree with those
determined by the spectral weight method, since the spectral
weight method is affected by the photoemission matrix elements.
When two features are very close and have similar intensities,
their spectral weight overlaps, and the positions of the local
spectral weight maxima shift. However, we stress that the spectral
weight method can give a qualitative and objective measurement of
the FS with the aid of the dispersion method, which relies on
subjective judgement.
\end{references}
\end{document}